# ACCELERATING CONTROL SYSTEMS WITH GITOPS: A PATH TO AUTOMATION AND RELIABILITY

M. Gonzalez*, M. Acosta†, Fermi National Accelerator Laboratory, Batavia, Illinois, USA

*Abstract*

GitOps is a foundational approach for modernizing infrastructure by leveraging Git as the single source of truth for declarative configurations. The poster explores how GitOps transforms traditional control system infrastructure, services and applications by enabling fully automated, auditable, and version-controlled infrastructure management. Cloud-native and containerized environments are shifting the ecosystem not only in the IT industry but also within the computational science field, as is the case of CERN and Diamond Light Source among other Accelerator/Science facilities which are slowly shifting towards modern software and infrastructure paradigms. The ACORN project, which aims to modernize Fermilab's control system infrastructure and software is implementing proven best-practices and cutting-edge technology standards including GitOps, containerization, infrastructure as code and modern data pipelines for control system data acquisition and the inclusion of AI/ML in our accelerator complex.

## INTRODUCTION

Fermilab's control system, ACNET [1], has supported accelerator operations for over four decades and continues to be the backbone of daily accelerator activities. Today, it comprises more than 600 active applications and services, developed primarily in FORTRAN, C++, and Java. These applications and services run on a mix of bare-metal servers and virtual machines, and most are accessed remotely through X11 over SSH. Source code has been maintained for decades in internal Concurrent Version System (CVS) repositories, reflecting the long-term continuity and stability of the system.

These practices were well suited to the technological context and operational needs of their time, providing robustness and reliability for accelerator operations. However, evolving requirements in scalability, collaboration, and security, along with advances in software engineering and infrastructure, now motivate a transition toward more modern approaches.

Several key shifts are being implemented by the Accelerator Control Operations Research Network(ACORN)[2] project as part of the modernization of Fermilab's control system:

- Architectural evolution: The system is moving away from monolithic applications that are tightly bound to specific hardware, toward microservices-based designs. This shift enables modularity, independent scaling of components, and faster development cycles.

---

* mariana@fnal.gov

† macosta@fnal.gov

- Ownership and collaboration: Instead of relying on a single expert or owner for each application, development is transitioning to team-based ownership. This model fosters peer reviews, improves resiliency of knowledge, and distributes responsibilities across multiple contributors.
- Programming paradigms: Legacy applications written in FORTRAN, C/C++, and Java are gradually giving way to more modern, web-based languages and frameworks.
- Security practices: The traditional approach of relying primarily on perimeter firewalls for protection is being replaced by systematic, standards-based security models. By following CISA and NIST hardening guidelines, the system adopts auditable, consistent, and adaptive practices that extend security beyond the network boundary.
- Infrastructure modernization: Applications and services are shifting from bare-metal and virtual machine deployments to containerized workloads orchestrated on Kubernetes[3] clusters. This provides declarative deployments, elasticity, automated recovery, and greater operational efficiency.

These changes create an environment where GitOps[4] emerges as a natural enabler, providing the necessary framework for automation, traceability, and reliability. By managing both infrastructure and applications as code, GitOps aligns with these modernization efforts and accelerates the integration of control systems into scalable and secure scientific environments.

## BACKGROUND

The ACNET control system comprises two main categories of applications and services: Console (CLIB framework) and Java [1]. Console applications are primarily written in C++ and are tied to the CLIB library; a monolithic framework that supports the integration of code and new functionality to control system software. The CLIB code base is the foundation for control system central services, data acquisition, data manipulation, network messaging and user-facing applications for several control system user groups. Applications that use CLIB as base are primarily written in C++ and use a tightly coupled monolithic system of shared libraries that enable machine experts, physicists and engineers to produce software that operates the Fermilab accelerator complex [1]. As a result, modifications and deployments to this framework and the applications that rely on it require careful peer-reviews and involve many manual steps that are not fully version-controlled or closely monitored through a streamlined pipeline.

The Management Environment for Controls Console Applications (MECCA) [5] is an in-house build system developed to automate compilation of the console applications. MECCA incrementally builds the application code base after changes are committed by developers. It is based on CVS version control and has been around for the majority of the lifetime of the ACNET control system.

In the early 2000s, several core applications and services were refactored in Java, shifting functionality from the C++ library (CLIB) into an object-oriented language. This move was motivated by the need to guarantee interoperability among various emerging Linux distributions and Windows consoles. One notable example of a core Java application is the Data Pool Manager (DPM), a central component of the control system that was re-implemented in Java. Over time, the DPM and other Java services and applications came to share a single source tree, which gradually evolved into another large monolithic structure, similar to the CLIB framework.

To automate the compilation and build for Java-based applications and services, a dedicated Java build system 'Codewatch' was introduced. It automates compilation, packaging, and release distribution of the Java tree. Built artifacts are written to shared storage (NFS), enabling multiple bare metal and VM nodes to run applications from the same release. However, because of extensive inter-dependencies across classes, isolating applications within this system has proved to be a major challenge. In addition, the use of CVS as version control for both the Console and Java code bases has made it difficult to implement automation and modern build system practices using CI/CD.

## KEY DEFINITIONS

### Infrastructure as Code

Infrastructure as code (IaC) is the ability to provision and support computing infrastructure using code instead of manual processes and settings [4], it uses code in declarative format to define the desired state of the system.

### Kubernetes

Kubernetes, also known as K8s, is an open source system for automating deployment, scaling, and management of containerized applications. It groups containers that make up an application into logical units for easy management and discovery [3]. Kubernetes operates by abstracting away a pool of compute resources and provides a uniform interface for workloads to be deployed and consume them. Its main component, the Controller, Works as an engine for resolving state by converging actual and the desired state of the system using Infrastructure as Code.

In a Kubernetes environment, Infrastructure as Code enables developers and operators to define, version, and manage Kubernetes resources using declarative languages like YAML or JSON, in complete alignment with GitOps' declarative nature.

### Kubernetes Operator

Kubernetes implements the Operator pattern, where a software extension makes use of Custom Resource Definitions (CRDs or extensions of the Kubernetes API [6]) to manage applications and components in an automated way. A Controller software component implements the operator logic by watching CRDs and reconciling components with the desired state declared by the user.

### GitOps

GitOps is a way of managing infrastructure and applications so that whole system is described declaratively and version controlled in a Git repository, and having an automated process that ensures that the deployed environment matches the state specified in a repository. [9]

### Flux

Flux CD is an open-source continuous delivery and GitOps tool designed to simplify and automate the deployment and lifecycle management of applications and infrastructure on Kubernetes [9]. With Flux CD, developers, and operators can declaratively define the desired state of their applications and configurations as code stored in a Git repository.

Flux CD continuously monitors the repository for changes and automatically applies updates to the Kubernetes cluster, ensuring that the actual state matches the desired state. By adopting the GitOps approach, Flux CD enables teams to achieve a reliable and auditable deployment process while promoting collaboration and traceability across different environments. [10]

## FROM SOURCE CONTROL TO AUTOMATED DEPLOYMENT

The adoption of GitOps in control systems follows a declarative, version-controlled, and automated workflow that enables both infrastructure and applications to be reproducible and traceable. The proposed methodology builds on three foundational principles:

- Git as the single source of truth: all configurations for applications, infrastructure, and testing environments are stored in Git repositories. Any change must be proposed via pull requests, ensuring traceability and peer review.
- Declarative configuration: system state is defined using Kubernetes manifests and Kustomize overlays, making deployments consistent and portable across environments.
- Automated reconciliation: FluxCD continuously monitors the repositories and ensures that the deployed state in the Kubernetes cluster matches the declared state in Git.

Figure 1 illustrates the proposed CI/CD pipeline as implemented in the GitOps workflow. The process begins with developers working on feature branches and submitting Pull Requests (PRs) for peer review. Automated CI pipelines are triggered on PR submission, running static code analysis, builds, and unit tests. Upon approval, container images are generated, published to a secure registry, and synchronized with the declarative GitOps repository. FluxCD then reconciles the declared and actual state within Kubernetes clusters, ensuring consistency and traceability.

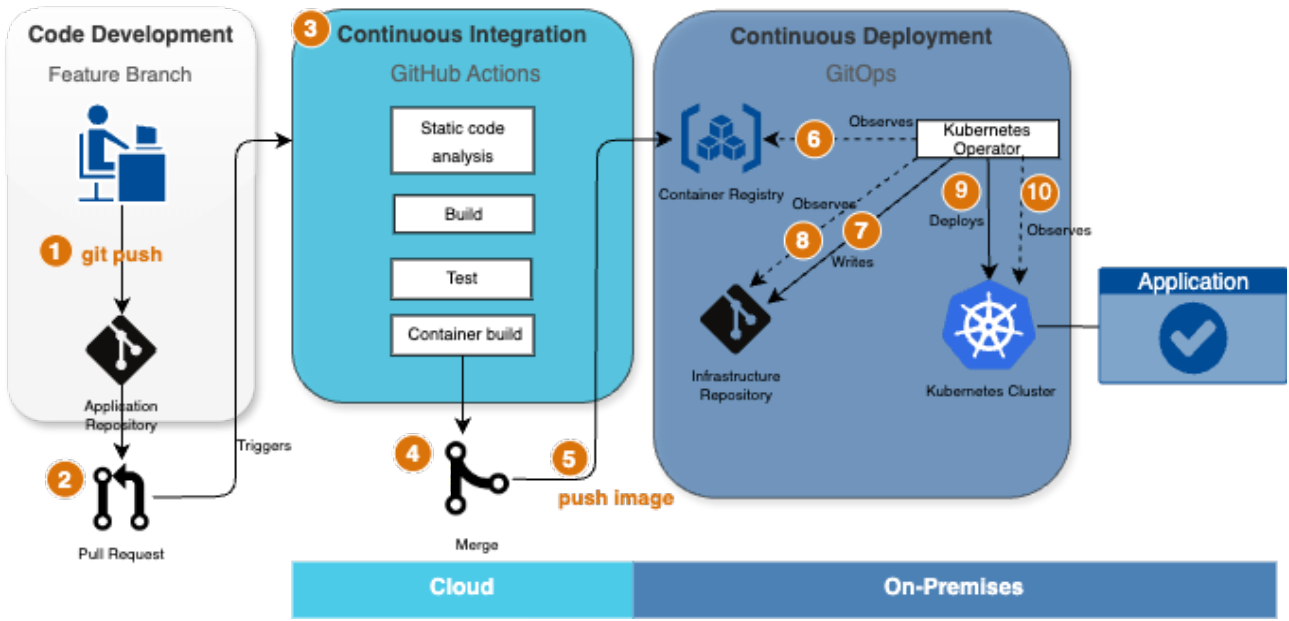

Figure 1: Continuous Integration and GitOps-based Continuous Deployment pipeline for control systems.

### Deployment process

The end-to-end CI/CD pipeline in the GitOps framework follows a nine-step process:

1. Engineer creates feature branch.
2. Pull Request (PR) submission triggers validation.
3. CI pipeline executes static code analysis, build, unit tests, and review.
4. Approved changes are merged, and a container image is built.
5. The image is published to a secured container registry with RBAC and semantic versioning.
6. The GitOps operator detects new image versions.
7. IaC repository is updated to record the system state.
8. Deployment to Kubernetes cluster is synchronized.
9. Continuous reconciliation ensures system consistency and drift remediation.

### IMPLEMENTATION

As part of the ACORN initiative, a custom Ansible playbook was developed to automate the installation and deployment of Flux. This is the very first step executed after

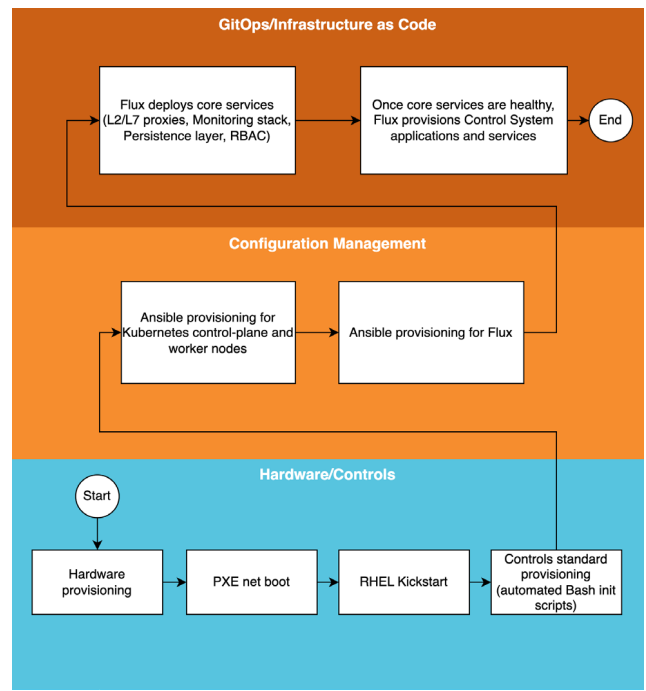

Figure 2: End-end Provisioning workflow

provisioning the Kubernetes control plane and networking components (CNI).

The rationale behind this approach is twofold:

- Infrastructure as Code for core services: all essential components—including proxies, monitoring, storage, and RBAC policies—are deployed declaratively with Git as the single source of truth, and Flux as the reconciliation engine.

- Lowering the barrier for developers: by abstracting the underlying infrastructure complexity, developers interact only with declarative manifests and Git workflows, rather than manual cluster management.

The process of creating a new cluster has been reduced to a four-step workflow:

1. Provision hardware.
2. Install Flux using the Ansible playbook.
3. Allow Flux to automatically deploy all core services and policies.
4. Operate the cluster without manual interventions.

This workflow significantly reduces manual setup effort, accelerates cluster on-boarding, and ensures reproducibility across environments.

## RESULTS

The GitOps-based CI/CD pipeline has been adopted for the development of web-based applications and the deployment of EPICS services supporting the Proton Improvement Plan II (PIP-II) project. These applications, unlike the legacy ACNET console and Java applications, are designed with containerization and declarative deployments from the ground up.

By integrating the pipeline, developers benefit from:
- Automated validation and deployment: each Pull Request triggers static code analysis, builds, and tests, ensuring quality before code is merged.
- Containerization and reproducibility: all artifacts are built into versioned container images, guaranteeing consistent runtime environments.
- GitOps-driven deployments: Flux automatically propagates changes to Kubernetes clusters, maintaining synchronization between declared and actual states.

### A successful prototype use-case: DPM

As part of ongoing control system modernization efforts, the Data Pool Manager (DPM) has been isolated from the larger monolith and its source code migrated it to GitHub. It has also been decoupled from the legacy Java Build System and a new build pipeline using maven and GitHub Actions has been put in place. Deployment of binaries is now managed with Ansible, enabling more agile updates, maintainability and the inclusion of container builds into the pipeline.

A test instance of the Data Pool Manager (DPM) is now running on a Kubernetes cluster managed by FluxCD, fully integrated into the GitOps workflow. This demonstrates that a legacy central control system service can be successfully containerized, version-controlled, and automatically reconciled, providing reproducibility, consistency, and reliability in a production-like environment. As part of this workflow, ACORN has leveraged Kubernetes liveness and readiness probes as means to ensure service reliability. A HTTPS '/status' endpoint was developed for DPM where each instance publishes a general health status as well as basic metrics in Prometheus format.

The Kubernetes-based DPM is in a production-ready state and it has been capable of servicing high speed, real control system Data Acquisition requests while maintaining uptime and optimal response times. Future work for this project includes stress testing, chaos engineering and the incorporation of more robust tests within the CI/CD pipeline.

ACORN has set ambitious goals for rapid application development, aiming to deliver new applications within two-week cycles. Early results indicate that GitOps and automated CI/CD pipelines significantly accelerate development, testing, and deployment, allowing teams to iterate quickly while maintaining reproducibility and reliability across the control system infrastructure.

This combined experience demonstrates the feasibility of applying modern software engineering practices to both legacy core services and new applications, providing a concrete pathway for gradually modernizing the Fermilab control system ecosystem.

## CONCLUSION

The adoption of GitOps, containerization, and Infrastructure as Code within the ACORN project demonstrates a practical pathway for modernizing Fermilab's legacy ACNET control system. By leveraging automated CI/CD pipelines, declarative deployments, and Flux-driven reconciliation, the project has enabled reproducible, reliable, and auditable workflows for both web-based applications and EPICS service deployments in the PIP-II project.

While these practices are common in modern software engineering, their application in a 40-year-old control system with hundreds of active applications illustrates the feasibility and benefits of bringing scientific infrastructure into a cloud-native, automated, and maintainable paradigm. This approach not only accelerates development and deployment but also provides a foundation for scaling, reproducibility, and long-term maintainability across the accelerator control ecosystem.

## ACKNOWLEDGMENTS

This manuscript has been authored by FermiForward Discovery Group, LLC under Contract No. 89243024CSC000002 with the U.S. Department of Energy, Office of Science, Office of High Energy Physics.

The prototyping and implementation of the present work would not have been possible without the participation of many people within Fermilab's Controls Department and both the ACORN and PIP-II projects. We thank everyone for their effort and apologize to anyone we missed in the following list: Charlie King, Usharani Kandula, John DeVoy, Jim Smedinghoff, Rich Neswold, John Diamond, William Badgett, Pierrick Hanlet, Tim Zingelman, Beau Harrison, Denise Finstrom and Anthony Tiradani.

# REFERENCES

[1] K. Cahill *et al.*, "The Fermilab Accelerator Control System," *ICFA Beam Dynamics Newsletter*, no. 47, pp. 106–124, 2008. FERMILAB-PUB-08-605-AD.

[2] D. Finstrom, E. Gottschalk, "Introduction and Status of Fermilab's ACORN Project", in *Proc. ICALEPCS'23*, Cape Town, South Africa, Oct. 2023, pp. 401–403. doi:10.18429/JACoW-ICALEPCS2023-TUMBCMO20

[3] R. Voirin, T. Oulevey, and M. Vanden Eynden, "The State of Containerization in CERN Accelerator Controls", in *Proc. ICALEPCS'21*, Shanghai, China, Oct. 2021, pp. 829–834. 10.18429/JACoW-ICALEPCS2021-THBL03

[4] G. Knap, T. M. Cobb, Y. Moazzam, U. K. Pedersen, and C. J. Reynolds, "Kubernetes for EPICS IOCs", in *Proc. ICALEPCS'21*, Shanghai, China, Oct. 2021, pp. 835–838. 10.18429/JACoW-ICALEPCS2021-THBL04

[5] Kubernetes, https://kubernetes.io/

[6] Kubernetes Custom Resources, https://kubernetes.io/docs/concepts/extend-kubernetes/api-extension/custom-resources/

[7] AWS IaC, https://aws.amazon.com/what-is/iac/

[8] OpenGitOps, https://opengitops.dev/

[9] FluxCD, https://fluxcd.io/flux/concepts/

[10] CNCF FluxCD, https://www.cncf.io/blog/2023/09/15/what-is-flux-cd/

[11] A. Waller, "The MECCA Source Code Capture Utility", presented at *ICALEPCS 1995*, https://www-conf.kek.jp/ICALEPCS/icalepcs1995-cdrom/PDFS/TH1BC.PDF